\documentclass[9pt]{article}
\usepackage[margin=1in]{geometry}
\usepackage{booktabs}
\usepackage{authblk}
\usepackage{graphicx}
\usepackage[backend=biber, doi=true, natbib=true]{biblatex}
 \usepackage{amsmath}

\title{Vector Autoregression (VAR) of Longitudinal Sleep and Self-report Mood
Data}
\author{Jeff Brozena}
\affil{College of Information Sciences and Technology\\Penn State University, United States\\brozena@psu.edu}
\date{May 2, 2023}

\begin{document}

\maketitle

\begin{abstract}%

  Self-tracking is one of many behaviors involved in the long-term
  self-management of chronic illnesses. As consumer-grade wearable sensors have
  made the collection of health-related behaviors commonplace, the quality,
  volume, and availability of such data has dramatically improved. This
  exploratory longitudinal N-of-1 study quantitatively assesses four years of
  sleep data captured via the Oura Ring, a consumer-grade sleep tracking
  device, along with self-reported mood data logged using eMood Tracker for
  iOS. After assessing the data for stationarity and computing the appropriate
  lag-length selection, a vector autoregressive (VAR) model was fit along with
  Granger causality tests to assess causal mechanisms within this multivariate
  time series. Oura's nightly sleep quality score was shown to Granger-cause
  presence of depressed and anxious moods using a VAR(2) model. 

\end{abstract}

\section{Introduction}\label{introduction}

Long-term self-management of chronic illnesses such as bipolar disorder require
persistent awareness of illness state over long periods of time and at varying
time scales
\citep{murnane2016SelfmonitoringPracticesAttitudes,morton2018TakingBackReins,majid2022ExploringSelftrackingPractices}.
Remaining consistently aware of key indicators signalling the onset of a
chronic condition allow individuals a chance at early intervention to reduce
the severity of a givene episode. For example, an individual may modify
behavior, engage their health practitioners, or adjust medication dosage.
However, bipolar disorder is an illness that often degrades an individual's
self-awareness and capacity for self-monitoring during symptomatic periods.

In the context of this specific illness, a volume of prior work has
demonstrated the vital role of sleep in order to promote mood stability and
prevent symptomatic episodes
\citep{harveySleep2009,murrayCircadian2010,gruberSleep2011}. Although the
particulars of this topic fall beyond the scope of this paper, these nuanced
relationships may in fact be self-reinforcing and bidirectional --- poor sleep
may lead to episodic onset, which may also lead to worsening (or shortening)
sleep bouts.

Given the importance of sleep in the ongoing management of this illness,
accurate consumer-grade alternatives to polysomnography (considered the gold
standard of sleep tracking) have emerged over the last few years. Indeed,
comparatively inexpensive sleep tracking technologies like the Oura Ring have
dramatically improved the quality of information that can be used to augment
and inform these self-monitoring activities. Objective sensor-based tracking
technology can be complemented with subjective self-report measures in order to
form a more complete picture of physical and mental health across time. Given
the aforementioned interplay of sleep and mood, this combination of subjective
and objective tracking creates the possibility of longitudinal analysis --- and
potentially deepens one's capacity for self-awareness.

Following four years of consistent sleep and mood tracking, I sought to more
formally interpret the data I had collected to quantify what I had previously
intuited: that certain mood states could be understood (and potentially even
predicted) by recent sleep trends. Indeed, this intuition has been demonstrated
quantitatively in existing literature \citep{boseVector2017,
moshePredicting2021, jafarlouObjective2023}. As this work also demonstrates,
combining data from consumer wearable technology and subjective self-report
logs allows for a more comprehensive picture of health.

I will first describe the vector autoregression (VAR) method and subsequent
tests, namely the Granger causality test and an impulse response analysis, that
were performed to achieve these goals.

First, I will describe the methods used to achieve these goals, providing a
brief overview of vector autoregression, Granger causality, and impulse
response functions. Next, I will detail the findings of these methods on the
dataset. This work concludes with a discussion of the methods and their
potential applications in future work.

\section{Problem setup}\label{methods}

A multivariate time series analysis was performed using a vector autoregressive
(VAR) model fit using ordinary least squares. An optimal lag order was first
obtained using a combination of Akaike Information Criterion (AIC), Bayesian
Information Criterion (BIC), Hannan-Quinn Information Criterion (HQIC), and
final prediction error (FPE). After fitting a VAR(2) model on the multiple time
series data (outlined below), a Granger causality test was performed in order
to assess the predictive relationships between variables. Finally, an impulse
response analysis was plotted to further explore the temporal relationships
between variables, specifically between sleep and self-reported mood.

\subsection{Vector Autoregression}\label{vector-autoregression}

A VAR(\textit{p}) model for a multivariate time series is a regression model
for outcomes at time \textit{t} and time lagged predictors, with \textit{p}
indicating the lag. Given \textit{p} = 1, the model would be concerned with one
observation prior to \textit{t}. As noted by \citet{lutkepohlNew2005} (as cited
in \citealt{seabold2010statsmodels}), a \(T \times K\) multivariate time series
(where \(T\) is the number of observations and \(K\) is the number of
variables) can be modeled using a \textit{p}-lag VAR model, notated as

\begin{equation} \begin{split} Y_t = \nu + A_1 Y_{t-1} + \ldots + A_p Y_{t-p} +
u_t \\ u_t \sim {\sf Normal}(0, \Sigma_u) \end{split} \end{equation}

where \(A_i\) is a \(K \times K\) coefficient matrix.

Intercept terms are included in \(\nu\) and regression coefficients are
included as the subscripted \(A\) values. This equation is solved using
ordinary least squares (OLS) estimation. The vector autoregressive (VAR) model
is a flexible method for the analysis of causality in this setting.

\subsection{Granger Causality Testing}\label{granger-causality}

Granger causality tests were performed in order to better assess the predictive
capacity of the Oura sleep score on self-reported mood states. Granger
causality defines one type of relationship between time series
\citep{grangerInvestigating1969} and states that a variable \textit{Granger
causes} another variable if ``the prediction of one time series is improved by
incorporating the knowledge of a second time series'' \citep{boseVector2017}.

Two autoregressive models are fit to the first time series, once with and
once without the inclusion of the second time series. The improvement of the
prediction is measured as the ratio of variance of the error terms. The null
hypothesis states that the first variable \textit{does not} Granger cause the
second variable and is rejected if the coefficients for the lagged values of
the first variable are significant.

Granger causation tests were applied using sleep scores as a single predictor
and each mood state as outcome variables.

\subsection{Impulse Response Function
Visualization}\label{impulse-response-analysis}

An impulse response function (IRF) is the ``reaction of a dynamic system in
response to an external change'' \citep{devriesWearableMeasured2023}. Plotting
an IRF allows for the interpretation of the impulse of a predictor on other
variables on subsequent periods. Given sleep scoring as a predictor, an IRF
visualization was created to better understand its impact on mood state. Figure
\ref{IRF} displays the results of this analysis over a 10-day period.

\section{Experimental Results}\label{results}

\subsection{Dataset Description}\label{dataset-description}

The sleep score dataset was created using the second- and third-generation Oura
Ring. The proprietary Oura sleep score is on a scale of 1 to 100 and
incorporates a variety of sensor-based measures (i.e., heart rate variability,
resting heart rate, body temperature) across time. Although the specifics of
this algorithm are not public, the Oura Ring has been found to produce accurate
measures of sleep timing and heart rate variability when compared against
polysomnography \citep{dezambottiSleep2019}. As detailed in Table \ref{Sleep},
my use of the Oura Ring was consistent across time. The dataset contains 1,455
nights of sleep bout data occurring between February, 2019 and March, 2023.

\begin{table}[hb] \centering \begin{tabular}{lr} \toprule & \textbf{Value}\\
\midrule Total nights & 1455 \\ Missing nights & 1 \\ Mean & 73.82 \\ SD &
12.36 \\ Max & 97.00 \\ Min & 30.00 \\ \bottomrule \end{tabular}
\caption{Descriptive statistics of Oura Ring sleep score data} \label{Sleep}
\end{table}

Each day at 4:30pm I received a notification prompting me to log my subjective
state in eMood Tracker, a mobile application for iOS. eMood Tracker is
``recommended by psychologists, therapists, and social workers'' and is
intended to ``track symptom data relating to Bipolar I and II disorders''
\citep{eMoods2023}. The version used through this period contains preset mood
categories (depressed, irritable, anxious, and elevated) and allow users to log
the presence and intensity on a scale of 0 to 3, where 0 is ``not present'' and
3 is ``severe''. The resulting dataset contains the most severe mood state per
day. The contents of this dataset are outlined in Table \ref{EMA}.

\begin{table} \centering \begin{tabular}{ll} \toprule \textbf{EMA Categories} &
\textbf{Count} \\ \midrule irritable & 100 \\ anxious & 88  \\ depressed & 103
\\ elevated & 48  \\ \bottomrule \end{tabular} \caption{Count of days where EMA
item contains a non-zero value} \label{EMA} \end{table}

\subsection{Data Analysis}\label{data-analysis}

All analysis were performed in Python version 3.11.0 \citep{Python} using
\texttt{pandas} 1.5.3 \citep{reback2020pandas} for data preprocessing and
\texttt{statsmodels} 0.13.5 \citep{seabold2010statsmodels} for modeling.
Dickey-Fuller tests of stationarity were performed using \texttt{statsmodels}
and the \texttt{pymdarima} library \citep{pmdarima}, a clone of R's
\texttt{auto.arima}.

\subsection{Stationarity, Decomposition, and
Autocorrelation}\label{stationarity-decomposition-and-autocorrelation}

A stationary time series contains no periodic fluctuation (``trend''). Without
stationarity, the means and correlations given by a model will not accurately
describe a time series' true signal \citep{boseVector2017}. If a time series is
found not to be stationary, an approach known as differencing can be applied to
achieve stationarity. The Dickey-Fuller test is one mechanism to determine
whether a time series is stationary.

Two Dickey-Fuller tests were performed on each time series, first via
\texttt{statsmodels} and then, additionally, using \texttt{pymdarima}'s
\texttt{should\_diff()} function to assess the need for differencing. The
\texttt{statsmodels} approach, an Augmented Dickey-Fuller test (ADF), yielded a
significant \textit{p}-value of .001 indicating support for the null hypothesis
that the time series is not stationary. However, the ADF performed via the
\texttt{pymdarima} approach using an alpha value of 0.05 yielded a
non-significant \textit{p}-value of 0.01 indicating that no differencing was
required in order to produce a stationary time series. For the purposes of this
study, I followed the results of the \texttt{pymdarima} library and assumed
stationarity.

\begin{figure}[h!] \centering
  \includegraphics[width=0.8\columnwidth]{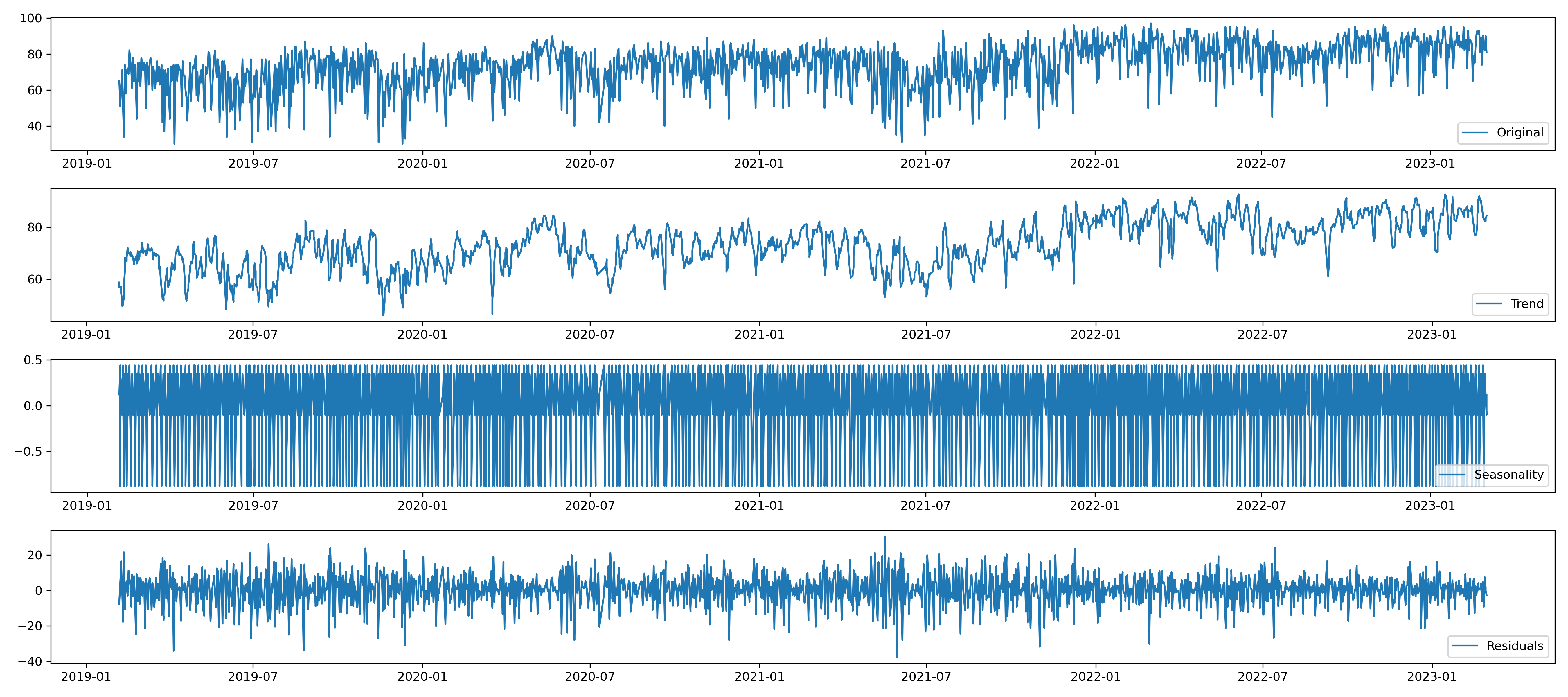}
  \caption{Decomposition of sleep time series} \label{Decomposition}
\end{figure}

\begin{figure}[h!] \centering
\includegraphics[scale=0.5]{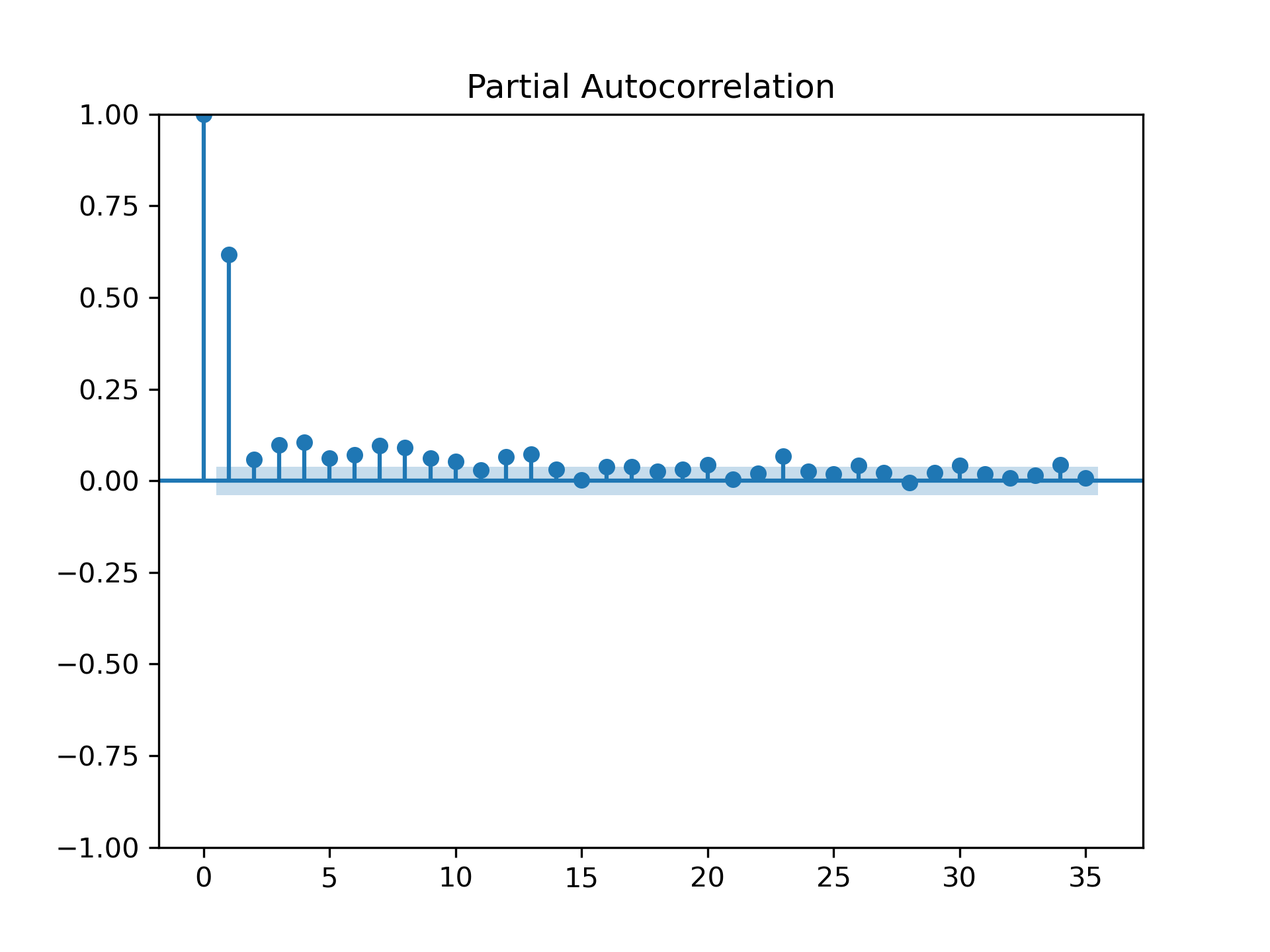} \caption{Partial
autocorrelation of sleep time series} \label{PACF} \end{figure}

An exploratory time series decomposition visualization was created to better
understand the presence of trend in the sleep score dataset. Figure
\ref{Decomposition} contains these results. Additionally, a partial
autocorrelation function was plotted using \texttt{statsmodels}, displayed in
Figure \ref{PACF}. Notably, partial autocorrelation appears to drop to zero for
lag values greater than 2.

\subsection{Lag Order Selection}\label{lag-order-selection}

The number of lags amount to the number of preceding days included as predictor
values in the model. The \texttt{statsmodels} function \texttt{select\_order()}
was used to assess an optimal lag value with possibilities between 0 and 15.
The optimal lag value was determined using four information criteria --- Akaike
Information Criteria (AIC), Bayes Information Criterion (BIC), Final Prediction
Error (FPE), and Hannan-Quinn (HQ) criterion. As detailed in Table
\ref{VAROrderSelection}, this selection process yielded a tie with lag-1 and
lag-2 each labeled as the minimum on these criteria. Rather than taking further
quantitative approaches, lag-2 was selected based on prior knowledge of sleep
quality and the onset of mood states.

\begin{table} \centering \begin{tabular}{lllll} \toprule ~ & \textbf{AIC} &
  \textbf{BIC} & \textbf{FPE} & \textbf{HQIC} \\ \midrule \textbf{0}  & 1.688
  &       1.708  &       5.408  &        1.695   \\ \textbf{1}  & -0.02482  &
  0.09412*  &      0.9755  &     0.01979*   \\ \textbf{2}  & -0.03545*  &
  0.1826  &     0.9652*  &      0.04635   \\ \textbf{3}  & -0.03417  &
  0.2830  &      0.9664  &      0.08481   \\ \textbf{4}  & -0.02907  &
  0.3872  &      0.9714  &       0.1271   \\ \textbf{5}  & -0.02537  &
  0.4900  &      0.9750  &       0.1680   \\ \textbf{6}  & -0.01701  &
  0.5975  &      0.9832  &       0.2135   \\ \textbf{7}  & -0.01940  &
  0.6943  &      0.9809  &       0.2483   \\ \textbf{8}  & -0.01014  &
  0.8026  &      0.9900  &       0.2948   \\ \textbf{9}  & -0.0008049  &
  0.9111  &      0.9993  &       0.3413   \\ \textbf{10} & 0.01201  &
  1.023  &       1.012  &       0.3913   \\ \textbf{11} & 0.02510  &
  1.135  &       1.026  &       0.4415   \\ \textbf{12} & 0.03723  &
  1.246  &       1.038  &       0.4909   \\ \textbf{13} & 0.04867  &
  1.357  &       1.050  &       0.5395   \\ \textbf{14} & 0.06022  &
1.468  &       1.063  &       0.5882   \\ \textbf{15} & 0.07076  &       1.577
&       1.074  &       0.6359   \\ \bottomrule \end{tabular} \caption{VAR Order
Selection (* highlights the minimum)} \label{VAROrderSelection} \end{table}

\subsection{Vector Autoregression Model}\label{vector-autoregression-model}

All time series under analysis were found to be stationary (ADF test \textit{p}
\textless .05). The results of the VAR(2) model predicting mood states using
sleep score are shown in Table \ref{VARScoreResults}. Sleep score was found to
be a significant positive predictor of depression, also confirmed via Granger
causation tests. Oura's sleep score did not positively or negatively predict other
mood states in this model.

\begin{table}[hb] \centering \begin{tabular}{lllll} \toprule ~ &
  \textbf{coefficient} & \textbf{std. error} & \textbf{t-stat} & \textbf{prob}
  \\ \midrule \textbf{L1.score} & 0.633262 & 0.027574 & 22.966 & 0.000 \\
  \textbf{L1.anxious} & 0.153275 & 0.446110 & 0.344 & 0.731 \\
  \textbf{L1.depressed} & 0.477164 & 0.409130 & 1.166 & 0.243 \\
  \textbf{L1.irritable} & -0.282988 & 0.412509 & -0.686 & 0.493 \\
  \textbf{L1.elevated} & -0.220198 & 0.655784 & -0.336 & 0.737 \\
  \textbf{L2.score} & -0.003080 & 0.027452 & -0.112 & 0.911 \\
  \textbf{L2.anxious} & 0.353528 & 0.445359 & 0.794 & 0.427 \\
  \textbf{L2.depressed} & 1.241873 & 0.409667 & 3.031 & 0.002 \\
\textbf{L2.irritable} & -0.080069 & 0.412341 & -0.194 & 0.846 \\
\textbf{L2.elevated} & -0.499540 & 0.657230 & -0.760 & 0.447 \\ \bottomrule
\end{tabular} \caption{VAR results for equation score} \label{VARScoreResults}
\end{table}

\subsection{Granger Causality}\label{granger-causality-1}

The results of the Granger causation tests are shown in Table \ref{Granger}.
Sleep score was shown to Granger-cause both depressed and anxious mood.

\begin{table}[h] \centering \begin{tabular}{llllll} \toprule \textbf{Causal
  Variable} & \textbf{Variable} & \textbf{Test statistic} & \textbf{Critical
  value} & \textbf{p-value} & \textbf{df} \\ \midrule sleepscore &
  \textbf{depressed} & \textbf{5.384} & 2.997 & \textbf{0.005} & (2, 6535) \\
  sleepscore & \textbf{anxious} & \textbf{3.294} & 2.997 & \textbf{0.037} & (2,
  6535) \\ sleepscore & irritable & 1.347 & 2.997 & 0.260 & (2, 6535) \\
  sleepscore & elevated & 1.203 & 2.997 & 0.500 & (2, 6535) \\ \bottomrule
\end{tabular} \caption{Granger Causality Tests for Sleep Score on Self-reported
Mood States} \label{Granger} \end{table}

\subsection{Impulse Response Analysis}\label{impulse-response-analysis-1}

\begin{figure}[h] \centering
\includegraphics[width=0.3\columnwidth]{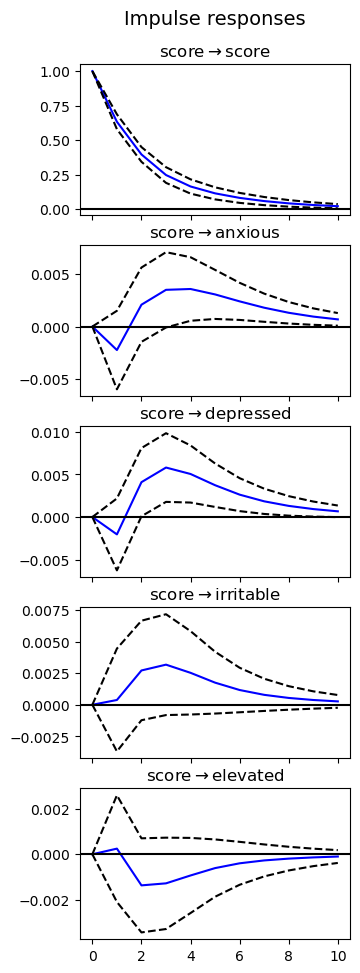} \caption{Plot of Impulse
Response Function, Lag 0 to 10} \label{IRF} \end{figure}

As shown in Figure \ref{IRF}, the impact of sleep score on the four
self-reported mood states varies over a 10-day period. Standard errors are
plotted at the 95\% significance level. The effect of an increase to sleep
score on depressed and anxious moods appear to be most felt only after several
days of its impact, peaking at roughly 3 days and then gradually decaying.

\section{Discussion}\label{discussion}

This exploratory study affirms the role that self-tracking technologies can
play in the ongoing management of affective disorders. This type of N-of-1
analysis would be impossible without inexpensive wearable sensors, and the
quality of this dataset is directly related to how non-invasive this particular
wearable is.

This work is not without limitation. My reliance on an algorithmic ADF test to
assess stationarity (rather than directly assessing the data myself) could
leave room for error. In the context of this work, an incorrect assessment of
stationarity risks the accuracy of the remainder of the analysis. Additionally,
this work only assesses the influence of the Oura sleep score on mood. In
reality, this is likely closer to a bidirectional influence and this should be
reflected properly in the analysis.

In future work, I hope to incorporate machine learning techniques for time
series segment annotation in order to explore the possibility of automatic
labeling of time periods where signals indicate the presence of an oncoming
episode.

\pagebreak

\printbibliography

\end{document}